\documentclass{article}

\usepackage{arxiv}
\usepackage{soul}

\usepackage[utf8]{inputenc} 
\usepackage[T1]{fontenc}    
\usepackage{hyperref}       
\usepackage{url}            
\usepackage{booktabs}       
\usepackage{amsfonts}       
\usepackage{nicefrac}       
\usepackage{microtype}      
\usepackage{siunitx}
\usepackage{graphicx}
\usepackage{subfig}
\usepackage[url=false,eprint=false]{biblatex}
\usepackage{doi}
\usepackage{xcolor}
\usepackage{amsmath}
\usepackage[capitalise,nameinlink]{cleveref}

\title{Beyond Active Engagement: The Significance of Lurkers in a Polarized Twitter Debate}

\author{
\hspace{1mm}Anees Baqir\\
	Ca' Foscari University of Venice\\
	Via Torino 155, 30172 Venice, Italy\\
	\And
\hspace{1mm}Yijing Chen \\
	Central European University\\
	Quellenstrasse 51, 1100 Wien, Austria\\
    \And
\hspace{1mm}Fernando Diaz-Diaz \\
	Institute for Cross-Disciplinary Physics\\and Complex Systems IFISC (UIB-CSIC)\\
	07122 Palma de Mallorca, Spain\\
    \And
\hspace{1mm}Sercan Kiyak \\
	Institute of Media Studies, KU Leuven\\
	Parkstraat 45 box 3603, 3000 Leuven, Belgium\\
    \And
\hspace{1mm}Thomas Louf \\
	Institute for Cross-Disciplinary Physics\\and Complex Systems IFISC (UIB-CSIC)\\
	07122 Palma de Mallorca, Spain\\
    \And
\hspace{1mm}Virginia Morini \\
	 University of Pisa,  Department of Computer Science\\
	Largo Bruno Pontecorvo, 3, 56127 Pisa PI, Italia\\
    \And
\hspace{1mm}Valentina Pansanella \\
	Scuola Normale Superiore\\
	P.za dei Cavalieri, 7, 56126 Pisa PI, Italia\\
    \And
\hspace{1mm}Maddalena Torricelli \\
	City University of London, Department of Mathematics\\
	London EC1V 0HB, United Kingdom \\
    \And
\hspace{1mm}Alessandro Galeazzi \\
	Ca’ Foscari University of Venice \\
	Via Torino 155, 30172 Venice, Italy\\
}

\date{\today}

\begin{document}
\maketitle

\begin{abstract}
The emergence of new public forums in the shape of online social media has introduced unprecedented challenges to public discourse, including polarization, misinformation, and the emergence of echo chambers. While existing research has extensively studied the behavior of active users within echo chambers, little attention has been given to the hidden audience, also known as lurkers, who passively consume content without actively engaging. This study aims to estimate the share of the hidden audience and investigate their interplay with the echo chamber effect. Using Twitter as a case study, we analyze a polarized political debate to understand the engagement patterns and factors influencing the hidden audience's presence. Our findings reveal a relevant fraction of users that consume content without active interaction, which underscores the importance of considering their presence in online debates. Notably, our results indicate that the engagement of the hidden audience is primarily influenced by factors such as the reliability of media sources mentioned in tweets rather than the ideological stance of the user that produced the content. These findings highlight the need for a comprehensive understanding of the hidden audience's role in online debates and how they may influence public opinion.
\end{abstract}


\section{Introduction}
The advent of the digital age has ushered in an era of unprecedented and instantaneous communication among members of society. 
While these technological advancements promised faster and wider access to information, their influence on the spread of information has turned out to be more nuanced. 
Indeed, they have also fostered several pervasive issues, such as polarization, misinformation, and the emergence of echo chambers that could influence public opinion and negatively impact society~\cite{andris_2015, neal_2020, falkenberg_2022}.
While these divergences could already be observed during the 20th century~\cite{neal_2020}, the introduction of social media networks may have increased the ideological divide among opposite factions~\cite{flamino_2023}. 
This radicalization in opinions has been shown to be a clear obstacle to dialogue, consensus, and policy-making~\cite{andris_2015, neal_2020}, being also considered as ``harmful to democracy and society''~\cite{polarization_harms_democracy} and a security risk for the UN~\cite{UNSafetySecurity2018}. 
Polarized debate is also a fertile environment for the spread of misinformation that may harm society at different levels~\cite{cinelli_2020b}. 
Falsehoods and unsubstantiated claims have been shown to spread widely in social media~\cite{vosoughi_2018, zhao_2020, juul_2021}, and they may erode trust in reliable sources \cite{lazer_2018}. 
One of the most insidious consequences of the digital age is the emergence of echo chambers \cite{cinelli_2021}, which have been found in various domains, including blogs \cite{gilbert_2009}, forums \cite{edwards_2013}, and prominent social networks \cite{cinelli_2021} like Facebook and Twitter \cite{del_vicario_2016b, cossard_2020}. 
While not intrinsically harmful, echo chambers may reinforce individuals' existing beliefs and perspectives, creating segregated environments where alternative viewpoints are suppressed and dissenting voices are silenced \cite{tokita_2021}. Moreover, the echo-chamber effect also exacerbates polarization and misinformation \cite{tornberg_2018}, trapping individuals within their own ideological bubbles and limiting the exploration of diverse perspectives.


In light of these pressing challenges, a surge of academic research has sparked over the last decade to understand the underlying mechanisms and real extent of echo chambers.
Scholars have dedicated substantial efforts to characterize them within online social networks systematically \cite{cinelli_2020, cinelli_2021, del_vicario_2016,morini2021toward,buongiovanni2022will}, and developed indices to gauge their presence and strength \cite{garimella_2018, diaz-diaz_2022, hohmann_2023}. 
Furthermore, various models have been proposed to elucidate the mechanisms driving the emergence of echo chambers \cite{baumann_2020, baumann_2021, diaz-diaz_2022}. 
While these models consistently emphasize the role of homophily as the primary catalyst for the echo chamber effect, a diverse range of contributing factors has also been proposed. 
Such factors include limited attention spans \cite{cinelli_2020}, selective exposure \cite{klapper_1960}, confirmation bias \cite{nickerson_1998}, the silencing effect \cite{tokita_2021}, and even the influence of feed algorithms \cite{brown2022echo}. 
Researchers have also explored methods to mitigate the impact of echo chambers, like introducing counter-biases within the feed algorithm \cite{vendeville_2023}. Nonetheless, it is important to note that some researchers contend that the influence of echo chambers may be overstated \cite{dubois_2018, barbera_2015, bruns_2017, de_francisci_2021}, thereby fostering ongoing debates surrounding the magnitude of their effects.

All the empirical studies mentioned above have one point in common: they focus on active users, meaning those who directly took action to interact with the content they were shown. On Twitter, for instance, those include liking a tweet, replying to it, or reposting it -- also known as retweeting.
But this might only be the tip of the iceberg: some users may actually belong to an echo chamber without actively participating in it.
These users are known as \emph{lurkers}. Although precise measurements of their relative prominence are not generally available, the current estimates place them as the majority of users on social networks, as they range from \SI{75}{\percent} to \SI{90}{\percent} \cite{GongCharacterizingSilent2015,AntelmiCharacterizingBehavioral2019}. Ignoring the presence of lurkers can lead to inaccurate estimations of echo chamber sizes and their potential impact on public debate. Consequently, an accurate estimation of their share is a pressing issue. 


In this work, we investigate the prominence of lurkers within the social network Twitter. To do so, we rely on the recently introduced metric called impression counts. An impression represents content appearing on a user's screen, reflecting visual engagement frequency. Notably, impressions quantify appearances, not unique viewers. Accordingly, this metric can be employed to estimate the share of the hidden audience, as well as the user engagement generated by different types of content. 
Additionally, we use this metric to explore whether the lurkers' share is influenced by factors like the ideological leaning of the content producer or the reliability and political bias of the sources used, thereby gaining insights into the lurkers' engagement patterns.

The remainder of this work is organized as follows: Section~\ref{sec:mat_and_met} describes the dataset we collect, alongside the methods and the analyses conducted on them. Section~\ref{sec:results} presents the main findings, and Section~\ref{sec:discussion} summarizes the strategy, highlights the results, and suggests future directions.

\section{Materials and methods}
\label{sec:mat_and_met}

\subsection{Data collection}
\label{sec:data}


We exploit two different datasets. The first aim is to investigate the interaction between ideological stances and user engagement, with a specific focus on assessing the share of hidden audiences in different communities of a polarized discussion. Accordingly, we collect through the official Twitter API all tweets related to the debate on whether countries should provide military support to Ukraine or not in the current 2023 war. This dataset consists of more than 17 million tweets posted by more than 5.2M users between November 22nd, 2022 and March 1st, 2023. The data collection process involved employing a search query that included the following terms: ``military aid'', ``military support'', ``tanks'', ``abrams'', ``leopard'', ``challenger'', ``jet'',  ``aircraft'', ``munitions'', ``HIMARS'', ``rockets'' and ``missile''. 



We use a second dataset assessing the political leaning and reliability of news outlets employed in this debate. This dataset is created starting from Media Bias/Fact Check (MBFC), an independent fact-checking organization that classifies news outlets based on their reliability and political bias - originally used in~\cite{cinelli_2021}. The dataset described above contains 2190 different news outlets, their domain names, political tendencies and reliability. The dataset was last updated in June 2019. These news outlets have been labeled according to their political leaning, ranging from "extreme left" to "extreme right." Additionally, some media sources are classified as "questionable" or "conspiracy-pseudoscience" if they have a tendency to publish misinformation or false content and endorse conspiracy theories.
To ensure a comprehensive analysis, we manually record the classification of these media outlets based on the information provided by MBFC, resulting in the inclusion of 468 outlets in addition to the existing pool of 1722 news outlets that already possess clear political labels. To calculate the individual leaning of users, each label is converted into a numerical value: -1 for Extreme Left, -0.66 for Left, -0.33 for Left-Center, 0 for Least Biased, 0.33 for Right-Center, 0.66 for Right, and +1 for Extreme Right, and the political leaning of a user is calculated as the average of the scores of all URLs it shared.


\subsection{Data filtering}
\label{sec:filter}


Considering our specific focus on engagement and hidden audience, we filter the dataset in order to maintain only the tweets whit a valid impression count. Firstly, we filter the dataset by date, retaining only the tweets posted after the introduction of the impression count metric, which is not available for tweets posted before the release of such metric (December 15th, 2022). Secondly, we restrict the dataset to tweets in English to avoid conflating factors (such as geography) that may affect the detection of users' ideological stances. 
Lastly, in the analysis of the hidden audience, we exclusively included original tweets. We disregarded other forms of content such as quotes, replies and retweets to accurately gauge the user's ability to engage their audience, as the impression on replies can be influenced by the original tweet and may not serve as a reliable proxy for measuring engagement.

\subsection{Interaction network}
\label{sec:network}

Using the dataset described in \ref{sec:filter}, we build the retweet interaction network. This methodology aligns with prevailing practices in Twitter analysis research~\cite{flamino_2023, falkenberg_2022,conover2011political}, as retweets are regarded as endorsements of content. On the other hand, quotes, retweets and replies are disregarded since they are less likely to signify endorsement and are often used for expressing opposing viewpoints or engaging in polemics~\cite{flamino_2023}. Using the English retweets dataset, we build a network by assigning a node to each unique user in the dataset - this includes users who either authored an original English tweet or retweeted an English tweet containing the specified keywords. We create directed edges from node A to node B if user A retweeted a post authored by user B and the weight of the edges is determined by the count of unique retweets between the two users, reflecting the strength of their interaction. The final interaction network counts 2.5 million nodes and 7.1 million edges. 

\subsection{Latent Ideology Estimation}
To estimate the ideological stance of users in the debate, we start from the latent ideology algorithm proposed in ~\cite{barbera_2015b, barbera_2015}. Following the studies already conducted in this field~\cite{flamino_2023, falkenberg_2022}, we consider retweets instead of follower/following relationships as interaction since retweets have been found to be good indicators of content endorsement~\cite{falkenberg_2022,flamino_2023}. The latent ideology algorithm requires the extraction of a subset of the influencer nodes which critically affects the ideology estimation results. The method by which such extraction is performed is the main topic of the following subsection. Once the influencer set is known, we apply the Correspondence Analysis algorithm \cite{greenacre2010correspondence}, which follows three steps: (i) Construction of the interaction matrix $A$, (ii) normalization of the matrix, and (iii) singular value decomposition. For the first step, we construct a matrix $A$, whose elements $A_{ij}$ represent the number of retweets user $i$ directs toward influencer $j$. Once $A$ is known, we normalize it as follows. First, we divide by the total number of retweets, obtaining:  

\begin{equation}
    P=\frac{A}{\sum_{ij} A_{ij}}. 
\end{equation}

Then, we define the following quantities: 

\begin{equation}
    \begin{cases}
    \textbf{r} = P \textbf{1}, \\
    \textbf{c} = \textbf{1}^T P, \\
    D_r = \text{diag}(\textbf{r}),\\
    D_c = \text{diag}(\textbf{c}),
    \end{cases}
\end{equation}

and we perform the following normalization operation: 

\begin{equation}
    S = D_r^{-1/2}(P- \textbf{r}\textbf{c}) D_c^{-1/2}
\end{equation}

For the third step, we perform a singular value decomposition of the form $S= U \Sigma V^T$, where $U, V$ are orthogonal matrices and $\Sigma$ is a diagonal matrix containing the singular values of $S$. Given the polarized nature of the networks under investigation, we can approximate the system by taking the subspace associated with the first singular value of the decomposition. Thus, we take the latent ideology of user $i$ to be the $i$-th entry of the first column of the orthogonal matrix $U$, while the median ideology of their retweeters represents the latent ideology of an influencer.

\subsection{Influencers Selection} 
As mentioned above, to apply the ideology-scoring algorithm we first need to extract a set of influencers from the retweet network. The influencer group encompasses several subgroups: (i) Russian and Ukrainian politicians, (ii) official accounts from information media sources such as journals and TV channels, and (iii) political activists.
Users in the retweet network are ranked according to their in-degree, according to the number of unique users who have retweeted them.
This enables us to start from a manually selected set of users pertaining to the three aforementioned categories with some of the highest in-degree.
This set then serves as a seed as we further select similar accounts using the "Who to follow" recommendations made by Twitter on their accounts' page.
We then refine the selection by excluding users with an in-degree lower than 100 and those whose content is unrelated to the Ukrainian conflict. These criteria yield a comprehensive set of 204 influencers.

\subsection{Estimation of the hidden audience proportion}
The estimation of hidden audience proportion leverages tweet-level metrics, including the number of likes, replies, retweets, quote tweets and, crucially, the number of impressions. We define the proportion of hidden audience as the ratio of the number of active actions taken -- namely, liking, replying, retweeting, and quote retweeting -- out of the number of impressions received by a given tweet, a given user, or a given domain, depending on the comparison unit of interest. We call this ratio \textit{Active Engagement (AE)}.
\begin{equation}
    AE = \frac{\text{\# of actions}}{\text{\# of impressions}}
\end{equation}
The measure of active engagement, coupled with users' popularity level, latent ideology, and the credibility and ideology of shared links, can give us crucial insights into how the share of the hidden audience may vary along these dimensions.


\section{Results}
\label{sec:results}
\subsection{Polarization in the debate around military support to Ukraine on Twitter}

\begin{figure}
    \centering
    \includegraphics[width=1\textwidth,trim={1cm 1cm 1cm 5cm},clip]{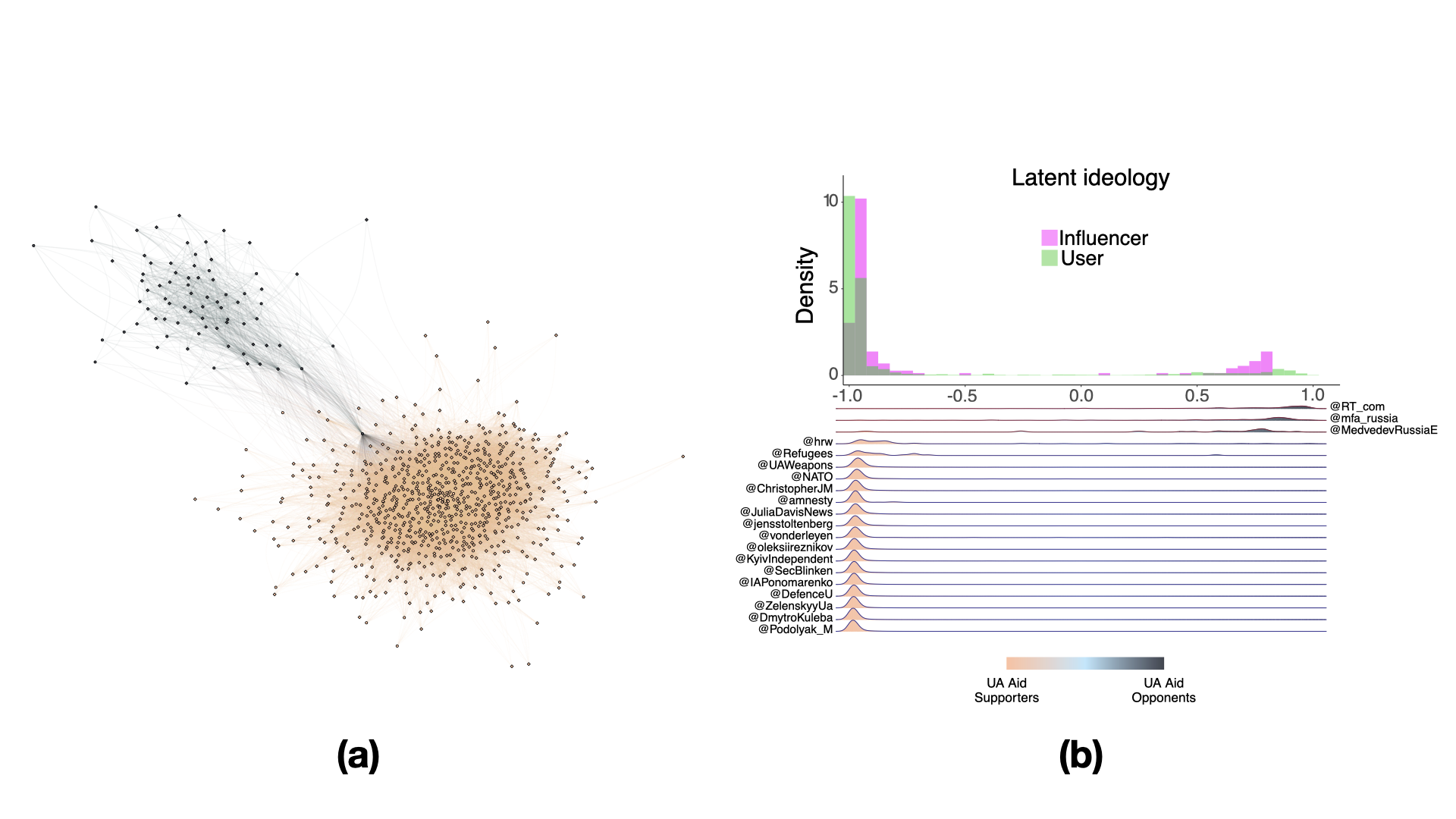}
     \caption{\textbf{Retweet Network and Ideology Distribution of users and influencers.} Panel (a): influencers and users retweet networks for nodes with a degree greater than 100, with edges colored on the base of nodes' ideological stances. Panel (b): histogram of users' and influencers' ideology score (top) and distributions of top influencers' retweeters' ideology obtained with the latent ideology algorithm (lower). Negative values represent pro-military aid alignment, while positive values correspond to military aid opponents. Bar colors in the top panel of (b) represent the density of influencers (pink color) and users (green color). Consistently with the color palette in panel (a), the area below retweeter distributions in the bottom panel of (b) is shaded in salmon if the influencer was inferred to be a supporter of Ukrainian aid, and in black if the influencer is against providing weapons to Ukraine.}
    \label{fig:ideology}
\end{figure}

The ongoing debate on whether other countries should provide military assistance to Ukraine during its conflict with Russia has generated significant attention from influential figures such as politicians, journalists, and committed citizens. As discussed above, the formation of echo chambers, where users predominantly interact with like-minded peers, is a common phenomenon observed in such controversial debates within social networks. Examining the presence of echo chambers around this polarizing topic is our first step to analyzing the dependence of hidden audience share on ideological stance. We estimated users' stances by computing ideology scores based on the influencers they had retweeted. Our results demonstrate a highly polarized discussion, with individuals advocating for military aid to Ukraine disproportionately engaging with like-minded users and those against military intervention engaging mainly with others holding similar views, as shown in panel a of \cref{fig:ideology}. The emergence of two distinct user clusters where they primarily retweet ideologically congruent influencers indicates a fairly limited volume of cross-ideology influence (see \cref{fig:ideology}(a)). Also, the latent ideology analysis shows a clear bimodal distribution of the users' and influencers' opinions as shown in \cref{fig:ideology}(b), further highlighting the presence of echo chambers in this debate (see also~\cref{fig:echochambers}).

\subsection{Unveiling the hidden audience in the echo chambers}

Having identified two opposing echo chambers, we now turn to the characterization of the hidden audience of this debate and how they may be distributed between these two groups.





\begin{table}[htbp]
  \centering
  \begin{tabular}{lll}
    \toprule
    action & average AE (\%) & Pearson's r\\
    \midrule
    retweet & 0.2909   & -0.3469 \\
    reply  & 0.2479  & -0.5649 \\
    like & 1.1154 & -0.2250 \\
    quote & 0.0612 & -0.5690 \\
    \bottomrule
  \end{tabular}
  \vspace{0.2cm}
    \caption{\textbf{Summary statistics for the active engagement by kind of action taken.} We compute the average AE for all tweets (with and without actions), as well as Pearson's r for tweets with actions (with log-scaled values) to show its correlation with user popularity.}
    \label{tab:hidden_audience_user}
\end{table}

\begin{figure}[ht]
	\centering
    \includegraphics[width=1\textwidth,trim={1cm 1cm 0.7cm 5cm},clip]{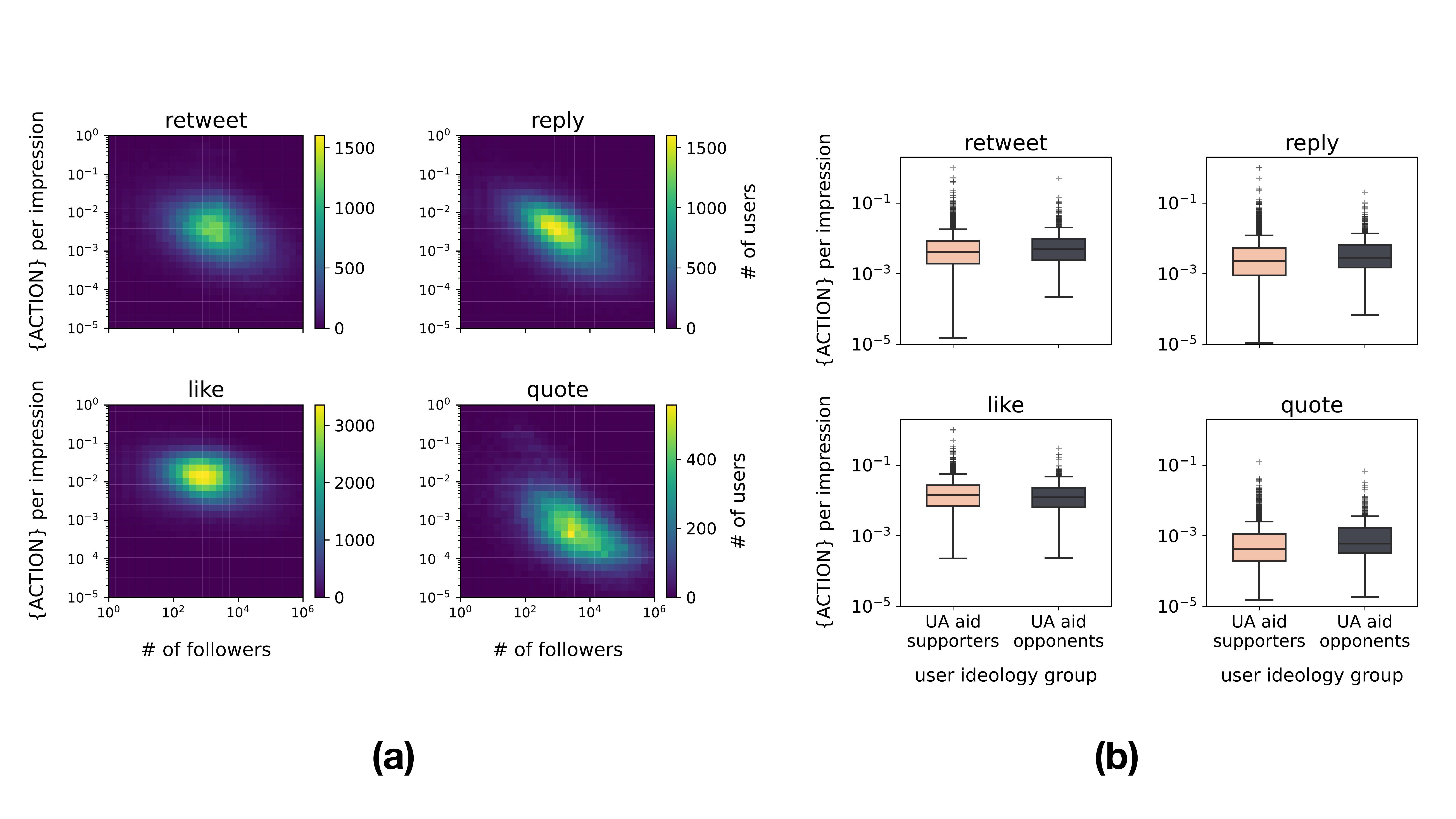}
        \caption{\textbf{User-level hidden audience in the Twitter discussion on military aid to Ukraine.} (a) Bivariate probability density of the number of followers and the active engagement with respect to retweets (top left), replies (top right), likes (bottom left), and quotes (bottom right). The active engagement seems to generally decrease with the number of followers of the original poster. (b) Boxplots of the active engagement for the same actions as in panel (a) grouped by users' ideologies, i.e., UA aid supporters (pink) and opponents (grey).}
	\label{fig:action_per_impression_user}
\end{figure}

\textbf{Across individual users.} We first show the proportion of hidden audience across individual users in \cref{fig:action_per_impression_user}(a). Overall, we see that the majority of the audience in such a polarized discussion is ``hidden'', as the average value of AE peaks at around 1\% and is particularly low for quote tweets (see \cref{tab:hidden_audience_user}). Out of the four actions identified in our data, the most prevalent action after viewing a tweet is liking (bottom left), the AE of which has the weakest correlation with the number of followers a user has. 
Such a weakly-correlated pattern extends to the AE of retweeting (top left), with a slightly higher level of average AE. On the other hand, when we look at actions that require textual inputs and may entail conversational interactions among users, such as replies (top right) and quote retweets (bottom right), there are more pronounced negative correlations between the AE level and the number of followers a user has (both log-scaled), and such a visual impression can be statistically verified by their Pearson's r values. 

\textbf{Across user groups.} Next, we turn our attention to the interplay between users' opinions and the hidden audience. Thus, we utilize the inferred ideology of individual users and compare AE levels between two opposing groups (i.e., UA aid supporters and opponents, see \cref{fig:action_per_impression_user}(b)). 
Across four actions, there are some subtle differences in these distributions, like a slightly higher AE for opponents' retweeting, replying, and quoting. That being said, we do not observe a qualitative difference overall, which means that these two groups, though opposed along the spectrum of ideology, do not appear to behave differently in terms of their engagement patterns.


\begin{figure}[!ht]
	\centering
	\includegraphics[width=.75\textwidth]{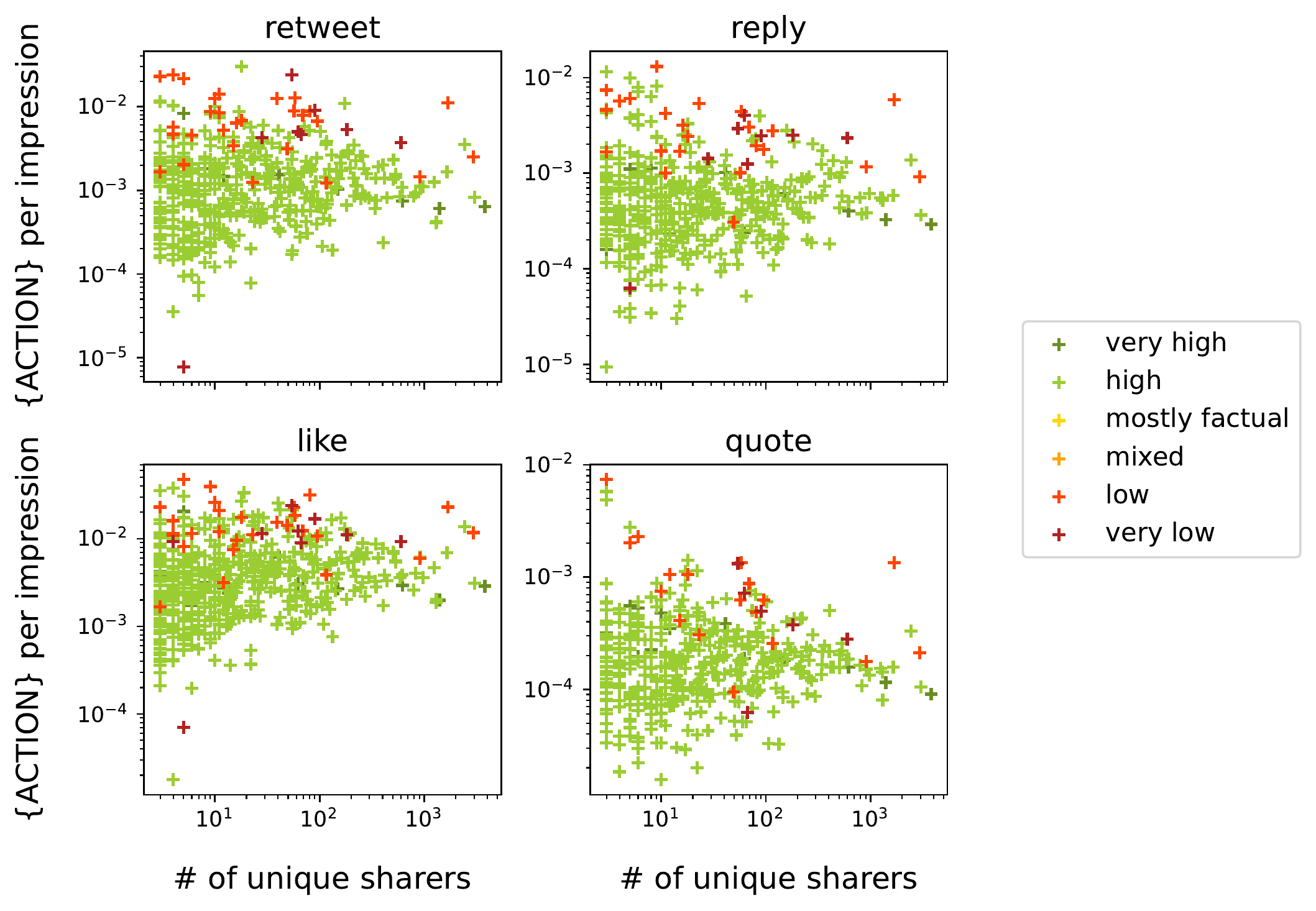}
	\caption{\textbf{Analysis of the hidden audience at the domain level.} Each scatter plot represents the relation between the number of unique sharers and the fraction of impressions that are followed by an action (either retweet, reply, like, or quote). Each cross represents a domain, and their color indicates their reporting quality.}
	\label{fig:action_per_impression_domain}
\end{figure}

\begin{figure}[!ht]
    \centering
    \includegraphics[width=.6\textwidth]{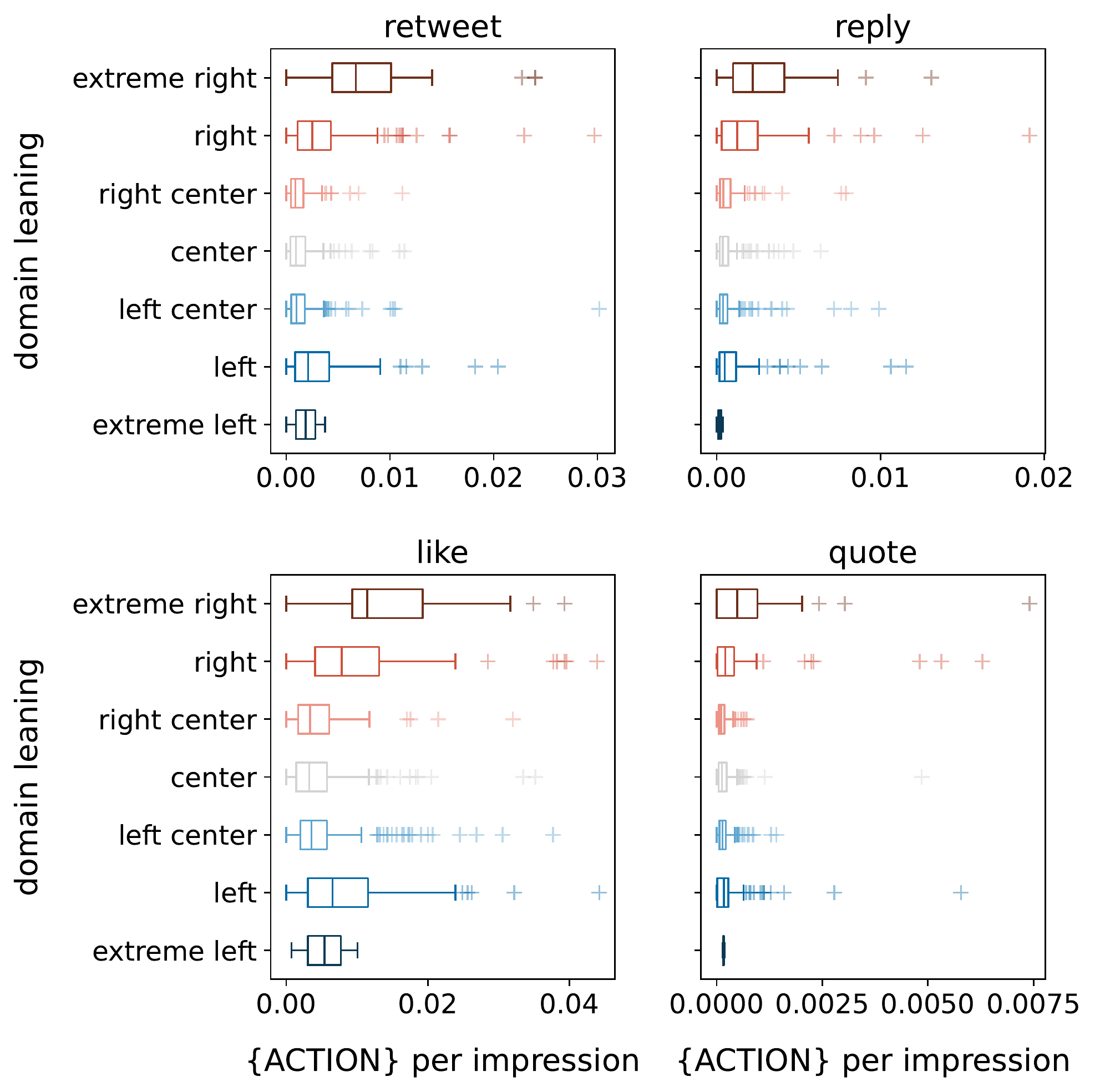}
    \caption{\textbf{Influence of political ideology on the share of the hidden audience.} Each subplot shows box plots representing the distribution of the fraction of impression that provokes an action (retweet, reply, like or quote). For each subplot, we group domains according to their political leaning into seven groups, ranging from extreme right to extreme left.}
    \label{fig:action_per_impression_domain2}
\end{figure}

\textbf{Across domains.} One possible factor influencing the hidden audience is the sources used in the tweets. 
Here we focus on news sources shared in the debate surrounding our topic of interest with respect to their reliability and political leaning, as shown in~\cref{fig:action_per_impression_domain,fig:action_per_impression_domain2}. Results show that the number of sharers does not have a strong influence on AE, as we observe across individual users. Similarly, the order of magnitude of AE does vary across different types of actions upon tweets sharing domain(s): the majority of likes maintain a relatively higher AE level around $10^{-3} \sim 10^{-2}$, while the AE of retweets and replies concentrates in a lower range of $10^{-4} \sim 10^{-3}$, with the AE of quotes being the lowest around $10^{-4}$.
, requiring more effort and direct interaction, 
Additionally, when analyzing the domains that predominantly report misleading information, we observe a consistently higher level of AE across four types of actions. This suggests that unreliable domains do not necessarily attract a larger number of unique sharers, but they do trigger more visible user-content interactions with smaller proportions of passive observers. Furthermore, among all reliability groups, the extreme-right domains have the highest AE level overall (\cref{fig:action_per_impression_domain2}), indicating that tweets sharing far-right domains engage a more actively engaged audience pool with a relatively smaller proportion of passive lurkers.

\begin{figure}[!ht]
    \centering
    \includegraphics[width=.6\textwidth]{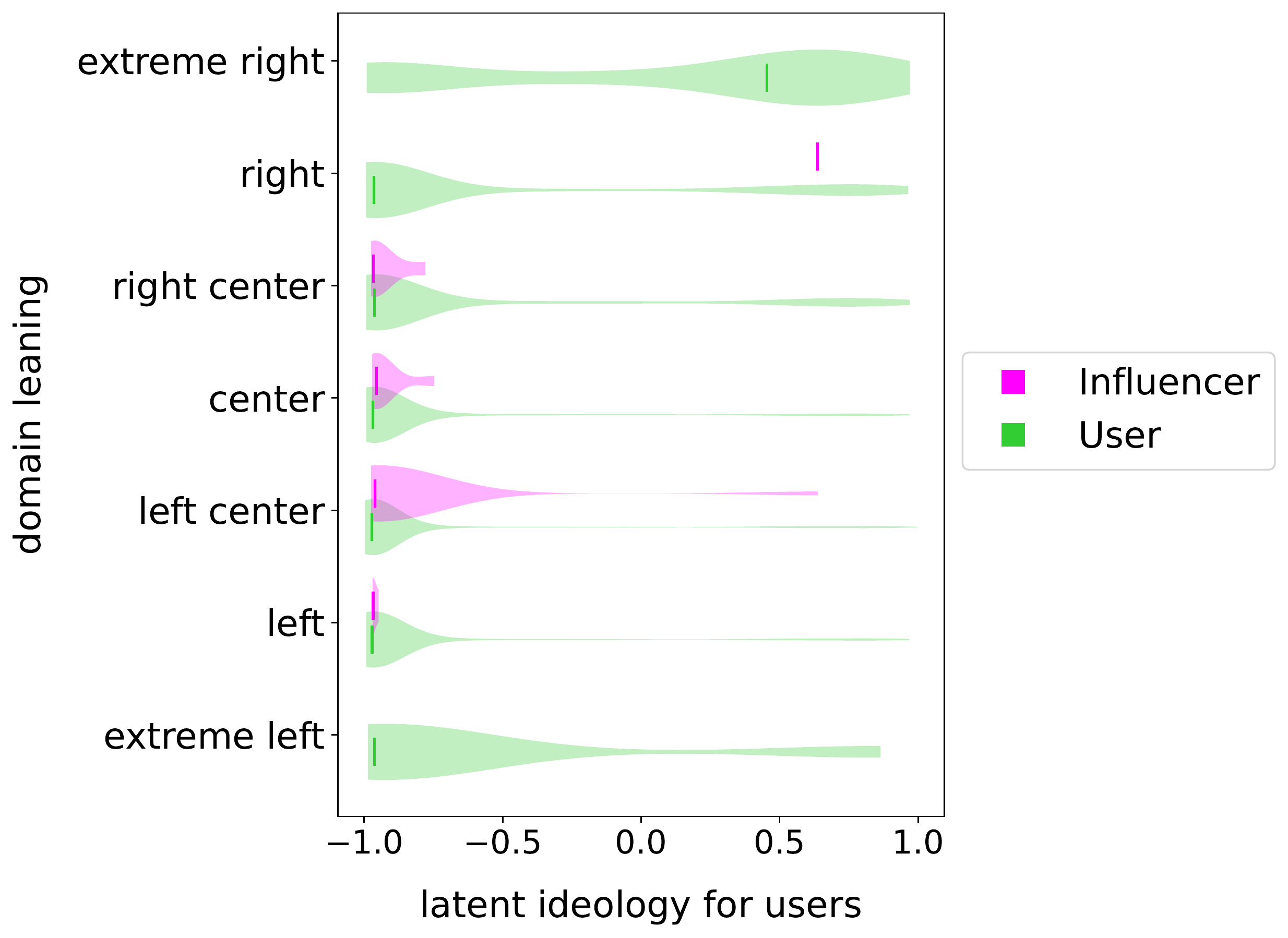}
    \caption{\textbf{Distributions of users by ideology segmented by political leaning of shared domains.} The distributions are shown for regular users and influencers who have shared at least a class of domains twice.}
    \label{fig:ideo_domain_validation_vp}
\end{figure}

As the next step, we combine the political leaning of domains shared by users and their detected positioning in the debate.
The ideology distribution of groups of users depending on the leaning of the domains they share is shown in \cref{fig:ideo_domain_validation_vp}.
Results highlight that support for the supply of weapons to Ukraine tends to drop the more extreme the political leaning.
However, there is still an important distinction between extreme left and extreme right, as the former does have a wider distribution, but the majority of its users are still in favor of military aid, while the latter opposes such aid in its majority.

\section{Discussion}
\label{sec:discussion}
This study presents a novel approach to investigating the interplay between echo chambers,  misinformation, and the share of users that consume content without visible interactions. The distinctive aspect of this approach lies in estimating the prominence of passive users, commonly referred to as lurkers, who refrain from actively engaging with tweets through actions such as liking, quoting, replying, or retweeting. 
Our work exploits the impression count, the newly introduced tweet-level metric from Twitter API, to estimate the proportion of the hidden audience that would be otherwise disregarded in analysis, inspiring a novel angle for analysis of polarization and echo chamber on Twitter, for which the previous analysis only take into account users with visible actions.

To address the presence of lurkers within the echo chamber, we compare the number of actions and impressions in the original tweets related to the Russo-Ukrainian conflict. This analysis allows us to determine the proportion of lurkers relative to active users and investigate their dependency on various factors. Our findings reveal that Twitter actions constitute a smaller portion compared to the total impressions, indicating that passive users account for a significant share of consumers. Furthermore, this share is even greater for actions that require active engagement, such as quoting. 
Notably, the main driver that influences the share of passive consumers is the presence of far-right and misinformation-spreading news sources. These contents exhibited the highest ratio of actions per impression, suggesting the dependence of active engagement on the type of content more than the ideological stance of the producers.

Although this study provides an initial understanding of the impact of passive users, it also raises several unresolved questions that warrant further investigations. Firstly, exploring whether lurkers are present in other debates of interest and social networks, such as Facebook or Reddit, would be valuable. However, we must consider that these platforms do not offer the same impression count metric yet. Secondly, the temporal variability of the action-per-impression rate remains unclear, necessitating a detailed analysis of lurker behavior over time. Thirdly, which types of media (e.g., URLs, videos, images) elicit the highest level of active user engagement is yet to be determined. Lastly, developing a minimal mechanistic model capable of reproducing the observed data would be instrumental in comprehending the underlying mechanisms driving content engagement. Furthermore, extending our exploration of lurkers to other debates and networks is crucial to gain a comprehensive understanding of their role in shaping online discourse.

In summary, this research makes a step forward by exploring the presence of passive users, who constitute a relevant part of social network users. Additionally, our findings underscore the heightened activity of users within domains with low factual reporting or far-right ideologies. By acknowledging the significance of lurkers and their relationship to echo chambers, polarization and misinformation, this study contributes to a comprehensive understanding of social network dynamics specifically related to the amplification of polarization on social media. Furthermore, it opens new avenues for future research and interventions aimed at addressing these challenges.

\section*{Acknowledgments}
This work is the output of the Complexity72h workshop, held at the IFISC in Palma, Spain, 26-30 June 2023. \url{https://www.complexity72h.com}

\printbibliography

\clearpage

\section*{Supplementary Information}
\label{sec:SI}

\setcounter{figure}{0}
\renewcommand{\thefigure}{S\arabic{figure}}

\begin{figure}[ht!]
    \centering
    \includegraphics[width=\textwidth]{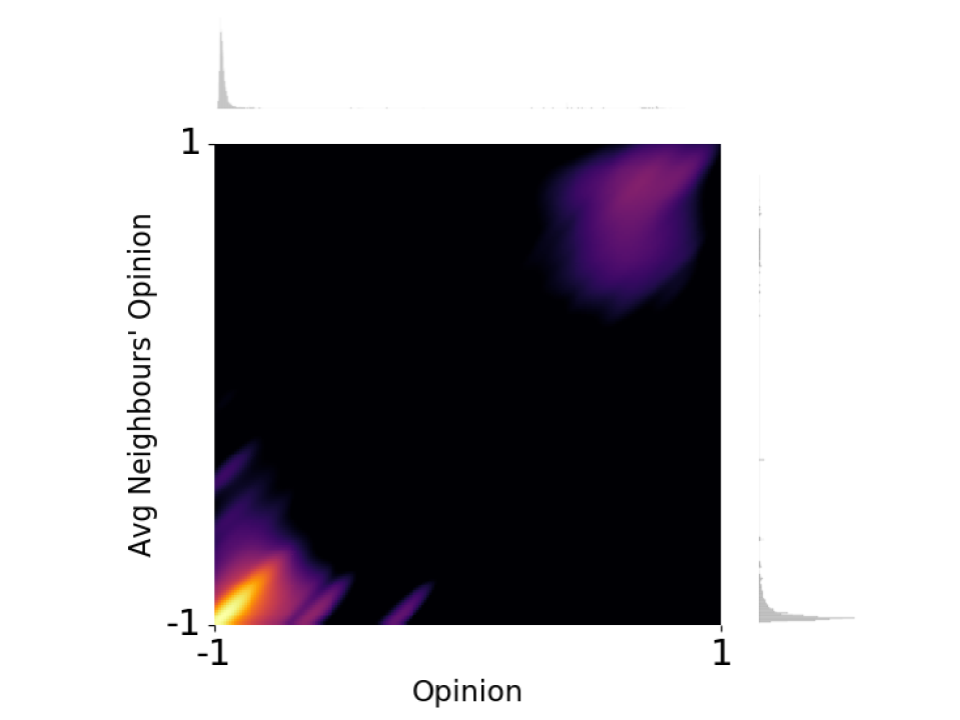}
    \caption{\textbf{Polarization and echo chambers in the interaction network.} The heatmap shows the users' opinion with respect to the average opinion of their neighbors; the colors represent the density of users: the lighter the color, the greater the number of users in that area. The values are computed in logarithmic scale in order to account for size differences between the two subpopulations. The heatmap clearly shows the presence of two subpopulations forming echo chambers, i.e., retweeting content posted by users with a similar ideology to theirs, in the considered discussion and time period. The density is computed on a logarithmic scale to account for such differences.}
    \label{fig:echochambers}
\end{figure}

We report here the additional figures and analyses discussed in the main text. Specifically, we show
in \cref{fig:echochambers} the probability density function of the bivariate distribution of the opinion of users (influencers excluded) and the average opinions of their neighbors to visualize the presence of echo chambers in the analyzed retweet network. The heatmap shows two opposing communities concentrated around the two extremes of the discussion, with different sizes. 

A similar analysis to the one shown in \cref{fig:ideology}(b) can be found in the data repository related to this project\footnote{\url{https://osf.io/5m3vr/}}. This figure shows the analysis conducted on all the 190 influencers of our dataset. The top panel of the figure represents the probability distribution of ideologies obtained using the latent ideology algorithm; the lower panel of the figure represents the opinion distribution of the retweeters of all influencers in the network. We hid the Twitter handles of the active users with fewer than 30,000 followers and labeled them as @Influencer$\_n$. In the same data repository, we also share the Twitter IDs of all the tweets used in this study.

\end{document}